\documentstyle[epsfig,aps,preprint]{revtex}
\preprint{FN-IEM/99/4, CTP\#2855}
\tightenlines
\draft
\begin{document}
\begin{titlepage}
\thispagestyle{empty}

\begin{center}

{\Large \bf Quasielastic Scattering from Relativistic Bound Nucleons:
Transverse-Longitudinal Response}

\vspace{0.5cm}
{\large J.M. Udias$^{1,a}$, J.A. Caballero$^{1,b}$, 
 E. Moya de Guerra$^{1}$,
J.E. Amaro$^{2}$ and T.W. Donnelly$^{3}$}

\vspace{.3cm}
{$^{1}$Instituto de Estructura de la Materia, CSIC 
Serrano 123, E-28006 Madrid, Spain}\\
{$^{a}$Departamento de F\'{\i}sica At\'omica, Molecular y Nuclear,
Universidad Complutense de Madrid, 
 E-28040 Madrid, Spain}\\
{$^{b}$Departamento de F\'{\i}sica At\'omica, Molecular y Nuclear,
Universidad de Sevilla, 
Apdo. 1065, E-41080 Sevilla, Spain}\\
{$^{2}$Departamento de F\'{\i}sica
Moderna, Universidad de Granada, 
E-18071 Granada, Spain}\\
{$^{3}$Center for Theoretical Physics,
 Laboratory for Nuclear Science and
Department of Physics, 
Massachusetts Institute of Technology, Cambridge, MA 02139, USA}\\

\end{center}

\vspace{0.5cm}
\begin{abstract}
Predictions for electron induced proton knockout from the $p_{1/2}$ and 
$p_{3/2}$ shells in $^{16}$O are presented using various approximations for 
the relativistic nucleonic current. Results for the differential cross section, 
transverse-longitudinal response ($R_{TL}$) and left-right asymmetry $A_{TL}$
are compared  at $|Q^2|=0.8$ (GeV/c)$^2$ corresponding to TJNAF experiment 
89-003. We show that there are important dynamical and kinematical 
relativistic effects which can be tested by experiment. 
\vspace{0.5cm}

\end{abstract}

\end{titlepage}

\newpage

\setcounter{page}{1}

 Under  quasielastic conditions the $(e,e'p)$ reaction
 can be treated with confidence \cite{Boffi}
 via the impulse aproximation in which 
 the detected proton absorbs the entire
 momentum ($\vec{q}$) and energy ($\omega$) lost by
  the electron. Detecting the outgoing proton in coincidence with
   the scattered electron allows one to determine
the missing energy ($E_m$) and momentum ($p_m$) in the
   reaction and thus to provide 
detailed information on the energies, momentum
 distributions and spectroscopic  factors of bound nucleons
  \cite{Boffi,Lapikas}.  Until
 recently low-$E_m$ data  were concentrated at $p_m\leq 300$ MeV/c. 
Now higher $p_m$-regions are
 being probed at small $E_m$ under  quasielastic conditions \cite{Irene,mainz}, 
yielding new information in a regime where 
two-body currents (meson exchange currents
 and $\Delta$-isobar contributions) can be safely neglected
  \cite{Ryckebush,ALC99}.

 Most theoretical work on $(e,e'p)$ has been carried out on the basis of
 nonrelativistic approximations to the nucleon current. Specifically, 
 the standard distorted wave impulse approximation (DWIA)  \cite{Boffi}
has been extensively employed \cite{Lapikas} to analyze $(e,e'p)$ data,
 using nonrelativistic current operators with bound and scattered 
proton wave functions deduced from
  phenomenological  nonrelativistic potentials.  However, DWIA  
 analyses have met two major difficulties:
  a)  The spectroscopic  factors extracted from low-$p_m$ data are too small 
compared with theoretical predictions \cite{Panda}.   For instance, 
 the extracted occupations of $3s_{1/2}$ and $2d_{5/2}$ orbits in $^{208}$Pb
 are $S_\alpha\simeq$ 0.5, while theories on short-range correlations 
 \cite{Panda} predict at most a 30\% reduction of mean-field occupations
 for levels just below the Fermi level.
  b) DWIA calculations compatible with the low-$p_m$ data predict much smaller
   cross sections at high-$p_m$
 than those  experimentally observed  \cite{Irene}.  Although
 short-range correlations are expected to increase the high-momentum
 components, their effect is negligible \cite{Muether}  at the small missing
 energies of these high-$p_m$ data, and effects of 
long-range correlations have been invoked \cite{Irene}. 

 In recent years the relativistic mean-field approximation
 has been sucessfully used for the analyses of both low-$p_m$
 \cite{Chi91,vanorden,Udi93,Udi95,otros} and high-$p_m$
  \cite{Udi96} data.
 In the relativistic distorted-wave impulse approximation (RDWIA), the nucleon
 current
\begin{equation}
J^{\mu}_{N}(\omega,\vec{q})=\int\/\/ d\vec{p}\/ 
\bar{\psi}_F
(\vec{p}+\vec{q}) \hat{J}^\mu_N(\omega,\vec{q}\/) \psi_B(\vec{p})
\label{nucc} 
\end{equation}
is calculated with relativistic   $\psi_B$ and $\psi_F$ 
 wave functions for  initial bound
and  final outgoing  nucleons, respectively.  
$\hat{J}^\mu_N$ is the
relativistic nucleon current operator of $cc1$ or $cc2$ forms
as in   \cite{deforest}.
 The bound state wave function is a four-spinor with well-defined parity and 
angular momentum quantum numbers and is 
obtained by solving the Dirac equation with scalar-vector
(S-V) potentials determined through a Hartree procedure from a relativistic
Lagrangian with scalar and vector meson terms \cite{SW}.
 The wave function for the outgoing proton is a  solution
 of the Dirac equation containing  S-V global optical potentials \cite{Hama}
 for a nucleon scattered with asymptotic momentum $\vec{p}_F$.
In contrast to DWIA, in RDWIA the analyses of individual nuclear shells
 were done with 
no fitting parameters other than the spectroscopic 
factors \cite{vanorden,Udi93,Udi95,otros}. 
The RDWIA  spectroscopic factors obtained are larger
than the DWIA ones \cite{Udi93,Udi95,otros} and are valid both for low- and 
high-$p_m$ data \cite{Udi96} --- for the above-mentioned
 $3s_{1/2}$ and $2d_{3/2}$ shells in $^{208}$Pb, values of
$S_\alpha\simeq 0.7$ have been obtained \cite{Udi93,otros} together with
reasonable agreement at high-$p_m$ \cite{Udi96}.

We have  recently studied  \cite{Cab98a,Cab98b}  the effect on the 
individual response functions
of the relativistic treatment of the nucleon current.
Focussed on the relativistic plane wave impulse approximation
(RPWIA) we showed \cite{Cab98a}
that the TL response is very 
sensitive to the  negative-energy components of the
 relativistic bound nucleon wave function.
We have also shown \cite{Cab98b} that,  for the $j=l\pm 1/2$ spin-orbit 
partners of a given shell, this sensitivity is
 much larger for the $j=l-1/2$ than for the $j=l+1/2$ case.
Strong sensitivity of the TL response to relativistic corrections
 was earlier found for $d(e,e'p)$ \cite{Gilad}.

A certain degree of controversy surrounds the TL 
response measured in  exclusive quasielastic electron  scattering 
from the least bound protons in several nuclei ($^{12}$C, $^{16}$O,
$^{208}$Pb): in some cases \cite{Spa93} large
deviations from standard DWIA calculations appear, while in others 
\cite{Chi91} the data are close to the calculations.
New data on the $R_{TL}$ response for proton knockout 
 from the $1p_{1/2}$ and $1p_{3/2}$ orbits  of $^{16}$O 
 are expected soon from an experiment
 carried out at Jefferson Laboratory (TJNAF)~\cite{Gao99}
 at
$|Q^2|\cong$ 0.8 (GeV/c)$^2$ and perpendicular
kinematics. The purpose of this work is to show for this case that 
there are important kinematical and dynamical relativistic effects.

For the bound state wave functions we use the parameters of the set NLSH
in \cite{Ring}. We have also obtained results with
the older HS set \cite{SW,Horowitz}, as well as with the newest NL3 one
\cite{newring}, and found very similar results.
For the scattered proton wave function,  we use the energy-dependent,
$A$-independent, potentials 
derived  by Clark {\em et al.} \cite{Hama} for $^{16}$O.
Again, we  checked that using other relativistic optical potentials
does not alter our results and conclusions.
The kinematical conditions follow closely those of the
TJNAF proposal \cite{Gao99}: beam energy 2445 MeV,
$|\vec{q}|\simeq 1$ GeV/c, $\omega\simeq439$ MeV,
corresponding to quasielastic kinematics
($\omega\simeq|Q^2|/2/M$), and proton kinetic energy of about 
$427$ MeV; the angle $\theta_F$ of the outgoing proton with respect
 to $\vec{q}$
is varied and thus the missing
momentum varies as $p^2_m=|\vec{q}|^2+|\vec{p}_F|^2-
2 |\vec{p}_F| |\vec{q}| \cos \theta_F$.

Figure~1 shows the differential cross section,
$R_{TL}$ response  and TL asymmetry ($A_{TL}$) for $p_{1/2}$ (top panels) 
and $p_{3/2}$ (bottom panels).  $R_{TL}$ and $A_{TL}$ are obtained from the
cross sections measured at $\phi_F=0^o$ and $\phi_F=180^0$ with
the other variables $(\omega,Q^2,E_m,p_m)$ held constant,
where $\phi_F$ is the azimuthal angle of the scattered proton (we follow 
the same convention as in \cite{Gao99}).
Relativistic calculations using the $cc1$ and $cc2$ 
current operators are shown by solid and dotted lines, respectively.
The Coulomb gauge has been used throughout and we have checked
that, as in our previous work (see in particular \cite{Cab98a}), the
Landau and Coulomb gauges produce similar results. On the contrary, the
Weyl gauge produces important deviations 
 and tends to give unrealistic enhancements of the longitudinal
current \cite{Cab98a}. We use a spectroscopic factor $S_\alpha=0.7$
as obtained for $^{208}$Pb \cite{Udi93}, which in RDWIA also matches 
the low-$p_m$ data \cite{Chi91,Leuschner} for the shells discussed here;
the spectroscopic factor simply scales down the curves for differential 
 cross sections and $R_{TL}$ while leaving $A_{TL}$ unchanged.

We can divide the differences between this fully relativistic approach and
the standard  nonrelativistic one into two categories:
$i)$ Effects due to the fully relativistic current
operator, {\em i.e.,} $4\times 4$ matrix structure of the 4-vector 
current operator, compared to the $2\times 2$ matrix structure of the
nonrelativistic current operator which is usually expanded in
$\vec{p}/M$. We call these effects {\em kinematical}
 as
they are in principle  independent of the dynamics introduced
by the nuclear interaction.
$ii)$ Effects due to the differences between relativistic and nonrelativistic
nucleon wave functions, which of course depend not only on the 4-spinor 
versus 2-spinor structure, but also on the potentials used in the 
respective Dirac and Schr\"odinger equations.
Thus we call these effects {\em dynamical}.

Further insight into these differences   can be gained recalling 
the steps taken in nonrelativistic approaches.
First, the one-body current operator is expanded in a
basis of free-nucleon plane-waves, which amounts to a truncation of the 
nucleon propagator that ignores negative-energy solutions
of the free Dirac equation. This is common to PWIA and DWIA. Second,
to take into account (spin-dependent) final-state interactions 
or distortion effects, in DWIA the current operator is transformed into a 
$2\times 2$ matrix (usually involving $\vec{p}/M$ expansions)
 to calculate the nucleon current as the matrix element
 between the nonrelativistic bound ($\chi_B$) and scattered
 ($\chi_F$) wave functions. This second step is not needed in
 PWIA, which can then better incorporate the relativistic kinematics,
 but misses important absorption effects. A ``relativized'' form
 of the $2\times 2$ current operator was proposed in 
 \cite{Ama96,Jes98} to optimize the $\vec{p}/M$ expansion.
 
 One may then identify two types of relativistic dynamical effects:
 
  $ii-a)$ Dynamical effects coming from the difference 
  between the upper components of $\psi_F$ ($\psi_B$) and the
  solutions $\chi_F$ ($\chi_B$) of the Schr\"odinger equation.
  Assuming equivalent central and spin-orbit potentials, this difference stems
  from the well-known Darwin term. Provided that the relativistic
  dynamics are known, one can deduce the Darwin term and construct an
  equivalent bi-spinor wave function $\chi$ to include its effect in the
  nonrelativistic nucleon current, thus removing this source of difference
  between relativistic and nonrelativistic results \cite{Udi95}. This is done
  for instance in ref. \cite{Kel97}. 
  The influence  of this term on $(e,e'p)$ observables has been demonstrated
  in several works 
\cite{Udi93,Udi95,Cannata,Jin94}. It appears to be the
main dynamical relativistic effect in the cross section in the
low-$p_m$ region  \cite{Udi93,Udi95}, and is important for the correct
determination of the spectroscopic factor from low-$p_m$ data. Its omission
reduces the spectroscopic factor by 15--20\%. It is included in all 
calculations presented here.

$ii-b)$ The other dynamical effect is due to the negative-energy 
components of the relativistic $\psi_B$, $\psi_F$ wave functions.
Starting from Schr\"odinger-like solutions $\chi$ one may at best
construct properly normalized four-spinors of the form
\begin{equation}
\psi=\frac{1}{\sqrt{N}}
\left(\chi(\vec{p}),\frac{\vec{\sigma}\cdot\vec{p}}{\bar{E}+M}\chi(\vec{p})
\right)
\label{eqx}
\end{equation}
to calculate the relativistic nucleon current. However,
this spinor lacks the dynamical enhancement of the lower component
of the Dirac solution due to the relativistic S-V potentials. This dynamical
enhancement is contained in the negative-energy components
of the relativistic $\psi_B$ ($\psi_F$) solutions and  influences 
 $(e,e'p)$ observables in the high-$p_m$ regions
 \cite{Udi95,Cab98a}.
   A discussion of how this effect may be incorporated
  in nonrelativistic formulations based on $2\times 2$ matrix,
   $\vec{p}/M$ expansions of current operators modified to include the
   effects of the relativistic S-V potentials can be found
   in \cite{Hedayati94}.


The role of the negative-energy components can be seen in
Fig.~1 comparing the solid with the short-dashed lines. 
The dashed lines show the results obtained
with the $cc1$ current operator when the negative-energy components
are projected out, {\em i.e.}, the nucleon 
current is calculated as
\begin{equation}
J^{\mu}_{N(++)}(\omega,\vec{q})=\int\/ d\vec{p}\/ \, 
\bar{\psi}_F^{(+)} 
(\vec{p}+\vec{q})
 \hat{J}^\mu(\omega,\vec{q})
\psi_B^{(+)}(\vec{p}),
\label{nuccproj} 
\end{equation}
where $\psi_B^{(+)}$ ($\psi_F^{(+)}$) is the positive-energy component
of $\psi_B$ ($\psi_F$), {\em i.e.},
\begin{equation}
\psi_B^{(+)}(\vec{p})=\Lambda_{(+)}(\vec{p}) \psi_B(\vec{p}),\,\,\,
\Lambda_{(+)}(\vec{p})=\frac{M+\overline{\slash\!\!\! p}}{2M} ,
\label{proj}
\end{equation}
with $\bar{p}_\mu=(\bar{E},\vec{p})$ and $\bar{E}=\sqrt{\vec{p}^2+M^2}$
(similarly for $\psi_F^{(+)}$).
The difference between the solid and short dashed lines is due to
the dynamical enhancement of the lower components which is contained
in the current of eq. (\ref{nucc}), but not in eq. (\ref{nuccproj}).
This effect is more visible than that introduced by the theoretical 
uncertainty due to the choice of $cc1$ (solid line) or $cc2$ (dotted line)
operators. It is important to realize that the positive-energy projectors
inserted in eq. (\ref{nuccproj}) depend on the integration
variable $\vec{p}$. One may attempt to neglect this dependence by using
projection operators corresponding to asymptotic values of
the momenta, {\em i.e.} projectors
acting on $\psi_F$ and $\psi_B$ respectively, with $P_F^\mu=(E_F,\vec{p}_F)$,
 $P_F^\mu-\bar{Q}^\mu$
the asymptotic four-momentum of the outgoing and bound nucleon respectively, 
with
$\bar{Q}_\mu=(\bar{\omega},\vec{q})$ and
$\bar{\omega}=E_F-\sqrt{(\vec{p}_F-\vec{q})^2+M^2}$. 
We refer to this approach as {\em asymptotic projection} ($J_{as}$). 
 The results corresponding to this
approximation are shown by long dashed lines in Fig.~1. They are obtained
with the $cc1$ operator and are very similar for $cc2$.

The differential cross sections for $|p_m|<300$ MeV/c are similar, 
but show a substantial dependence on the negative-energy 
components for $|p_m|> 300 $ MeV/c for either the $p_{1/2}$ or
$p_{3/2}$ shells.
Note also that
the cross sections obtained with positive-energy projected wave
functions are more symmetrical around $p_m=0$ than the
RDWIA results.
Therefore, the effect of removing the negative-energy components
shows up more in $R_{TL}$  and $A_{TL}$ (see middle and right-hand
panels of Fig.~1). Particularly interesting is the
oscillatory structure
of the fully relativistic  result for $A_{TL}$. This characteristic
 is preserved by the positive-energy projection method of
eq.~(\ref{nuccproj}), but not by the method of asymptotic projection. 
 Note that the dependence on the dynamical 
enhancement of the
lower components is stronger for the $p_{1/2}$ $R_{TL}$ response than for 
$p_{3/2}$, a feature that was first seen in RPWIA \cite{Cab98a} 
and that persists in RDWIA.
On the other hand, the asymptotic projection severely modifies 
$A_{TL}$ for both orbitals.
We notice that  the $A_{TL}$ calculated with $J_{as}$
approach are very similar to the ones obtained in \cite{Kel97}, as it
is similar to the  EMA (noSV) in said reference.
At low momentum this approach lies close to the fully relativistic one and
to the $J_{proy}$ ones,
but beyond $p_m\simeq 200$ MeV/c it gives completely different results. The 
oscillating trend of the  $A_{TL}$ calculated in RDWIA is confirmed
by the preliminary data \cite{Gao99} and agrees qualitatively with 
previous calculations by Van Orden \cite{Gao99}.

Other relativistic effects can be seen in Fig.~2, where we compare RDWIA
results on $A_{TL}$ (left panels) and $R_{TL}$ (right panels) to
nonrelativistic approaches at various levels. To minimize the differences
 we have used the $cc2$ current operator
 and nonrelativistic scattered wave functions obtained from Dirac-equivalent
 Schr\"odinger equations \cite{Udi95}. This ensures
 that the nonrelativistic wave functions correspond to the upper components
 of the relativistic ones, containing in particular the Darwin term. For the
 nonrelativistic bound wave functions, we used the ones in \cite{Ama96}.
 In \cite{Ama96,Jes98} new 
 approximations to the on-shell relativistic 
 one-body current operator were developed to take better account of 
 relativistic kinematic effects in nonrelativistic calculations. In particular,
  the charge density contains a spin-orbit correction that affects $R_{TL}$
  \cite{Ama96}. In Fig.~2 we show by long dashed lines the results obtained
  with the ``relativized current'' and by dotted lines the results obtained
  with that relativized current when the spin-orbit correction term to the
  charge-density is neglected. 
   One can see
  that the spin-orbit correction has a very large effect on $R_{TL}$ and
  $A_{TL}$. Its omission causes large deviations from the relativized
   current results. Using the DWEEPY \cite{DWEEPY} code we have obtained 
    similar results to the dotted lines in Fig.~2.

The short dashed lines in Fig.~2 are results obtained with the 
relativistic current   and 4-spinors constructed  as in
eq. (\ref{eqx}) from the 
nonrelativistic bound and scattered wave functions.
 In this way, the relativistic kinematics are fully
taken into account as well as the dynamical effect in the upper
component, namely the effect of the Darwin term; only the dynamical 
enhancement of the lower components is missed. This is why these results 
(short-dashed lines) for $A_{TL}$ and $R_{TL}$ 
 are much closer to the fully relativistic results
shown by the solid lines.
We see also that a large oscillation 
of $A_{TL}$ can be recovered  in the 
nonrelativistic approach. The much smaller oscillation of $A_{TL}$  
 is a distinctive feature of the
$J_{as}$ results, and it seems to be ruled out
by the experiment.
 We also note that, while the effect of the
dynamical enhancement of the lower components is larger in $p_{1/2}$ 
than in $p_{3/2}$ shells, the effects of relativistic kinematics are 
of the same order in both shells.

In conclusion, we have identified two types of relativistic effects on 
 $R_{TL}$ and  $A_{TL}$. One is of kinematical
origin, and has a large contribution from the spin-orbit correction
to the charge density, and other is of dynamical origin. The latter is
mainly due to the enhancement of the lower components and is stronger
for $p_{1/2}$ than for the  $p_{3/2}$ orbital.
 This is in addition to the dynamical effect on the upper component
 due to the Darwin term which is present in all the results
 given here, and that mostly affects the determination of spectroscopic
  factors\cite{Udi95}. It is encouraging that the preliminary data 
  \cite{Gao99} agree so well with
   the predictions of the fully relativistic calculations and one anticipates
  being able to make even more stringent tests when a finer grid of 
  high-precision data involving other nuclei
  become available in the range $200\le p_m \le 400$ MeV/c.

This work was partially supported under Contracts  No. 940183 
(NATO Collaborative Research grant), 
\#DE-FC01-94ER40818 (cooperative agreement with the 
US department of Energy D.O.E.),
 PB/95-0123,
PB/95-0533-A, PB/95-1204 (DGICYT, Spain), PB/96-0604
(DGES, Spain),
PR156/97 (Complutense University, Spain)
and  by the Junta de Andaluc\'{\i}a
(Spain).

\begin{figure}[t]
\begin{center}                                                                
\mbox{\epsfig{file=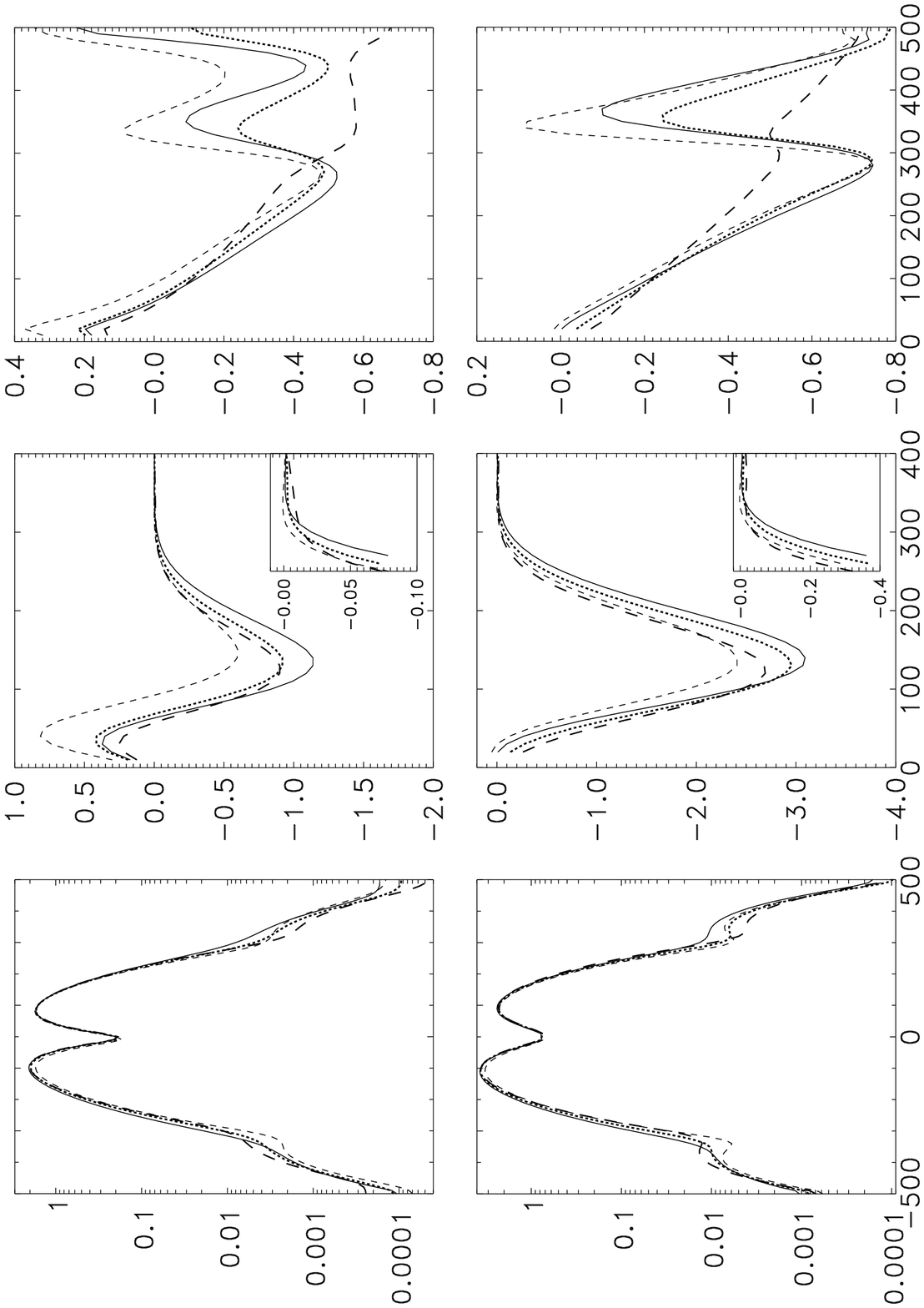,height=0.85\textheight}}
\end{center}
\caption{Cross section in nb/(MeV sr) (left), $R_{TL}$ 
in fm$^3$ (middle) and $A_{TL}$ (right) for proton knockout from $^{16}$O for
the $1p_{1/2}$ (upper panel) and $1p_{3/2}$ (lower panel)
orbits, versus missing momentum $p_m$ in MeV/c. Results shown 
correspond to a fully relativistic
calculation using the Coulomb gauge and the current 
operators: $cc1$ (solid line) 
and $cc2$ (dotted line). Also shown are the results after 
projecting the bound and scattered proton wave functions over positive-energy 
states ($J_{proy}$, dashed line) and using the asymptotic momenta 
($J_{as}$, long-dashed line). For the $p_{3/2}$ shell a small
contribution was taken into account from the nearby $5/2^+$ 
and $1/2+$ states known
from the low-$p_m$ data \protect\cite{Leuschner}.}
\end{figure}

\begin{figure}[t]
\mbox{\epsfig{file=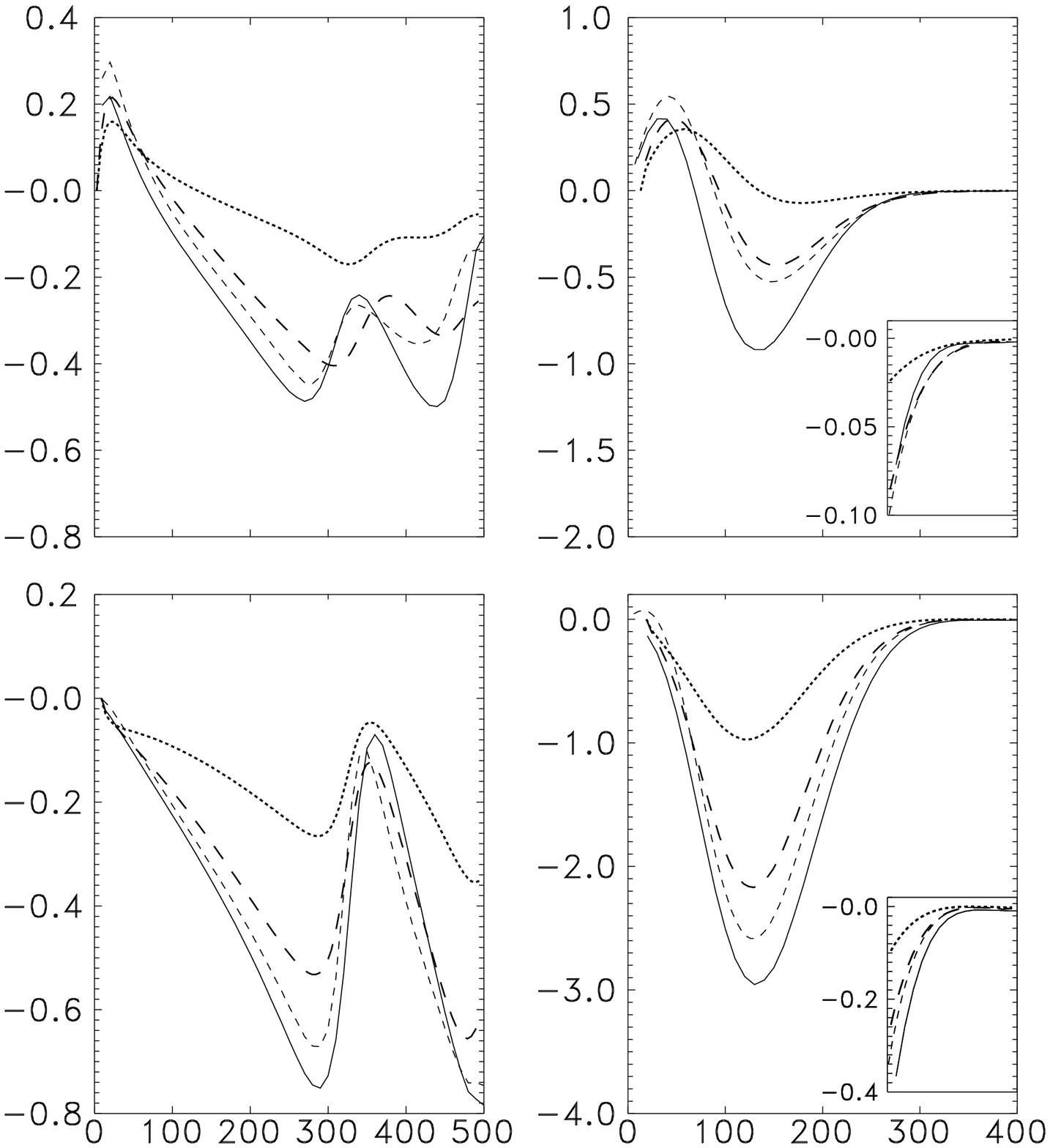,width=0.85\textwidth}}
\caption{$R_{TL}$ (right panels) and $A_{TL}$
asymmetries (left panels) for proton knockout from $^{16}$O for
the $1p_{1/2}$ (top panels) and $1p_{3/2}$ (bottom panels) orbits.
Results shown correspond to a fully relativistic
calculation using the Coulomb gauge and the current operator $cc2$ (solid line),
a calculation performed by
projecting the bound and scattered proton wave functions over positive-energy 
states (dashed line) and two nonrelativistic calculations with
(long-dashed) and without (dotted) the spin-orbit correction term in the
charge density operator (see text for details).}
\end{figure}

\end{document}